\documentclass{aa}
\usepackage{natbib,graphicx,hyperref}

\begin{document}
\title{Limits to solar cycle predictability: Cross-equatorial flux plumes}

\author{R.H. Cameron\inst{1} \and M. Dasi-Espuig\inst{1} 
\and J. Jiang\inst{2} \and E. I\c{s}{\i}k\inst{3} \and D. Schmitt\inst{1}
\and M. Sch\"ussler\inst{1}}

\institute{Max-Planck-Institut f\"ur Sonnensystemforschung, 
  37191 Katlenburg-Lindau, Germany
\and Key Laboratory of Solar Activity, National Astronomical Observatories,
Chinese Academy of Sciences, Beijing 100012, China
\and Department of Physics, Faculty of Science \& Letters, Istanbul K\"ult\"ur
University, Atak\"oy Campus, Bak\i rk\"oy 34156, Istanbul, Turkey}
\date{Received ; accepted}

\abstract
{Within the Babcock-Leighton framework for the solar dynamo, the strength of a 
cycle is expected to depend on the strength of the dipole moment or net hemispheric flux
during the preceding minimum, which depends on 
how much flux was present in each hemisphere 
at the start of the previous cycle and how much
net magnetic flux was transported across the equator during the cycle. 
Some of this transport is associated with the random walk of magnetic flux tubes subject 
to granular and supergranular buffeting, some of it is due to 
the advection caused by systematic cross-equatorial flows such as those 
associated with the inflows into active regions, and some crosses
the equator during the emergence process.}
{We aim to determine how much of the cross-equatorial transport is due to 
small-scale disorganized motions (treated as diffusion) 
compared with other processes such as emergence flux across the equator.}
{We measure the cross-equatorial flux transport using Kitt Peak synoptic 
magnetograms, estimating both the total and diffusive fluxes.}
{Occasionally a large sunspot group, with a large tilt angle emerges crossing the equator,
with flux from the two polarities in opposite hemispheres. The largest of 
these events carry a substantial amount of flux across the equator (compared to the magnetic flux 
near the poles). We call such events cross-equatorial flux plumes.
There are very few such large events during a cycle, which introduces an uncertainty
into the determination of the amount of magnetic flux transported across the equator in any particular cycle. 
As the amount of flux which crosses the equator determines the amount
of net flux in each hemisphere, it follows that the cross-equatorial plumes
introduce an uncertainty in the prediction of the net flux in each hemisphere. 
This leads to an uncertainty in predictions of the strength of the following cycle.}
{}
\authorrunning{Cameron et al.}
\titlerunning{}
\maketitle

\section{Introduction}
The large-scale field of the Sun, as measured at the surface, 
reverses roughly every 11 years. A signature of the reversal is the change of sign of the net flux 
in the southern and northern hemispheres, and the reversal of the polarity orientation of emerging bipoles 
in each hemisphere according to Hale's law.
Application of Stokes' theorem to the induction equation shows that the change in the 
net magnetic flux through the solar surface of the northern hemisphere is equal
to the amount of magnetic flux transported across the equator -- being the boundary of the
northern hemisphere surface \citep{Durrant04}.

On the Sun's surface the (molecular) diffusion is small, and the transport is almost entirely 
due to advection by either large-scale systematic flows (such as the Sun's differential rotation
and meridional circulation) or small-scale turbulent flows (such as are associated with granulation and 
supergranulation) -- the latter flows shuffle the field in random directions and can be treated 
as a diffusive process. 
The transport across the equator (at the surface) occurs either by horizontal transport of radial 
magnetic field across the equator, or by the radial transport 
of horizontal field through the solar surface (flux emergence) at the equator.

Cross-equatorial flux transport hence plays a critical role  in the 
evolution of the Sun's large-scale magnetic field \citep[for a discussion of
it in the context of solar cycle prediction see][]{Petrovay10}. 
In models such as the Surface Flux Transport model 
\citep[for a recent review, see][]{Mackay12}, 
or Flux Transport Dynamo model \citep{Choudhuri95}, 
the meridional flows are usually assumed to be anti-symmetric 
about the equatorial plane, and therefore do not transport flux across the 
equator. Exceptions in this regard include the work of \cite{Jiang10} and 
\cite{Cameron12a} who considered inflows into the active region belts,
which can occasionally extend across the equator. Flux can also emerge across the
equator, with different polarities in each hemisphere. The relative importance
of these different processes in transporting magnetic flux across the equator can be
assessed observationally.

In this paper we use the Kitt Peak synoptic magnetograms to 
estimate the total net flux transported across the equator. We evaluate this
as the time derivative of the net flux in the northern hemisphere.
We also estimate the amount which is transported by the small-scale
random motions (i.e. the amount of magnetic flux transported across the 
equator by turbulent diffusion). The difference between the two
is the amount of flux transported by processes which can not
be described using the diffusive approximation. This non-diffusive  
component is shown to be associated 
with discrete events in which highly tilted bipolar groups emerge across 
the equator, or near the equator and some of the flux is then advected 
across the equator. In this paper we study examples of each type of event.

\section{Examples of cross-equatorial transport }
\begin{figure*}
\begin{center}
\includegraphics[scale=0.95]{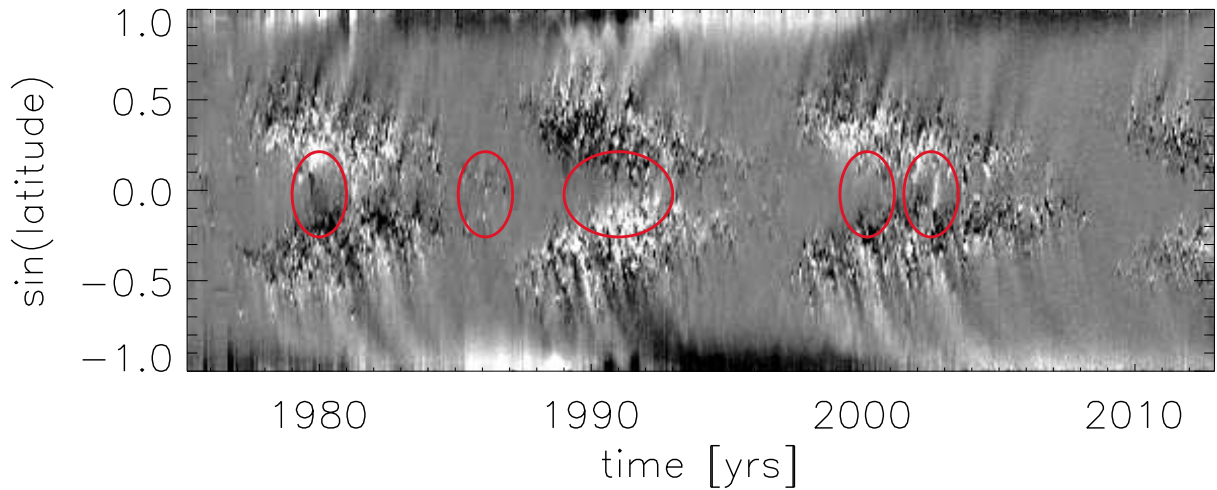}
\caption{Longitudinally averaged radial magnetic field as a function of latitude 
and time from the Kitt Peak Solar Observatory synoptic magnetograms. The 
grey scale is saturated at $\pm 10$G. The red circles illustrate 
cross-equatorial flux plumes. Black indicates negative magnetic field, white postive.}
\label{fig:mbf}
\end{center}
\end{figure*}

Our analysis is based on the National Solar Observatory, Kitt Peak synoptic maps of 
the radial component 
of the Sun's magnetic field, consisting of one map per Carrington rotation
\citep{Harvey98}.
The maps for Carrington rotations 1625 to 2007 are based on the NSO Vacuum
Telescope\footnote{The NSO Vacuum Telescope data were obtained from 
\href{ftp://nsokp.nso.edu/kpvt/synoptic/}{ftp://nsokp.nso.edu/kpvt/synoptic/}},
thereafter we used the synoptic maps based on the SOLIS 
telescope\footnote{The SOLIS data was obtained from 
\href{ftp://solis.nso.edu/synoptic/level3/vsm/merged/carr-rot/}{ftp://solis.nso.edu/synoptic/level3/vsm/merged/carr-rot/}}. 

\subsection{Cross-Equatorial Flux Plumes}
The data gives the radial component of the magnetic field strength
as a function latitude, $\lambda$, longitude, $\phi$, and time, $t$. 
Hereafter we consider the longitudinally averaged field
$B_r(\lambda,t)$, which is shown in Figure~\ref{fig:mbf}.

Notable features include the wings of the butterfly  diagram and 
the inclined features which extend from the active regions towards the poles.
These features were studied by  \cite{Durrant01} and called 'flux plumes'.
Although they are fewer  
in number, similar inclined features can be seen crossing the equator. 
The red ellipses in Figure~\ref{fig:mbf} outline the larger of these
events. We call these
`cross-equatorial flux plumes', owing to their similarity with the 
`flux plumes'.

\subsection{Emergence across the equator}
\begin{figure}[h!]
\begin{center}
\includegraphics[scale=0.45]{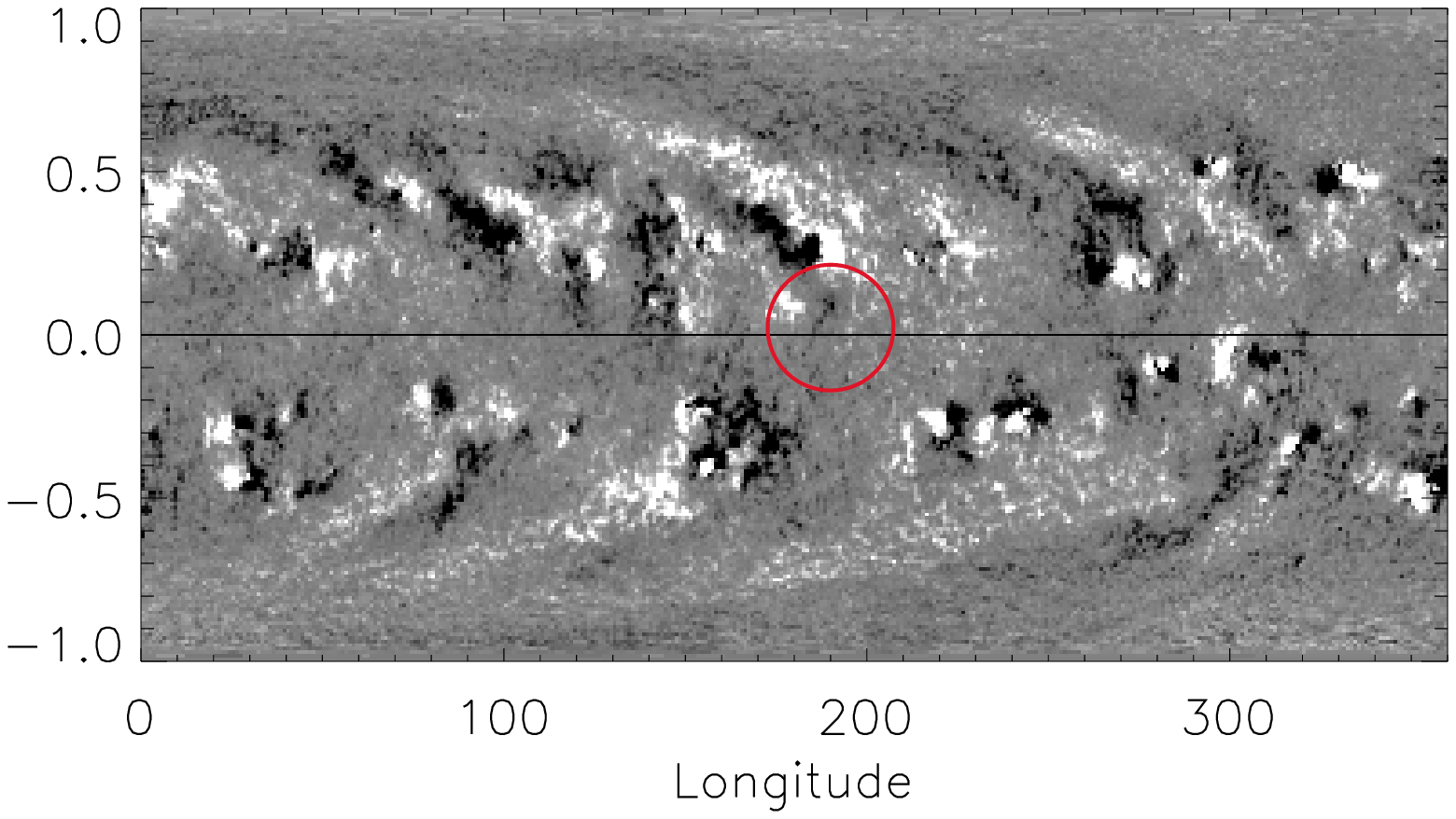}
\includegraphics[scale=0.45]{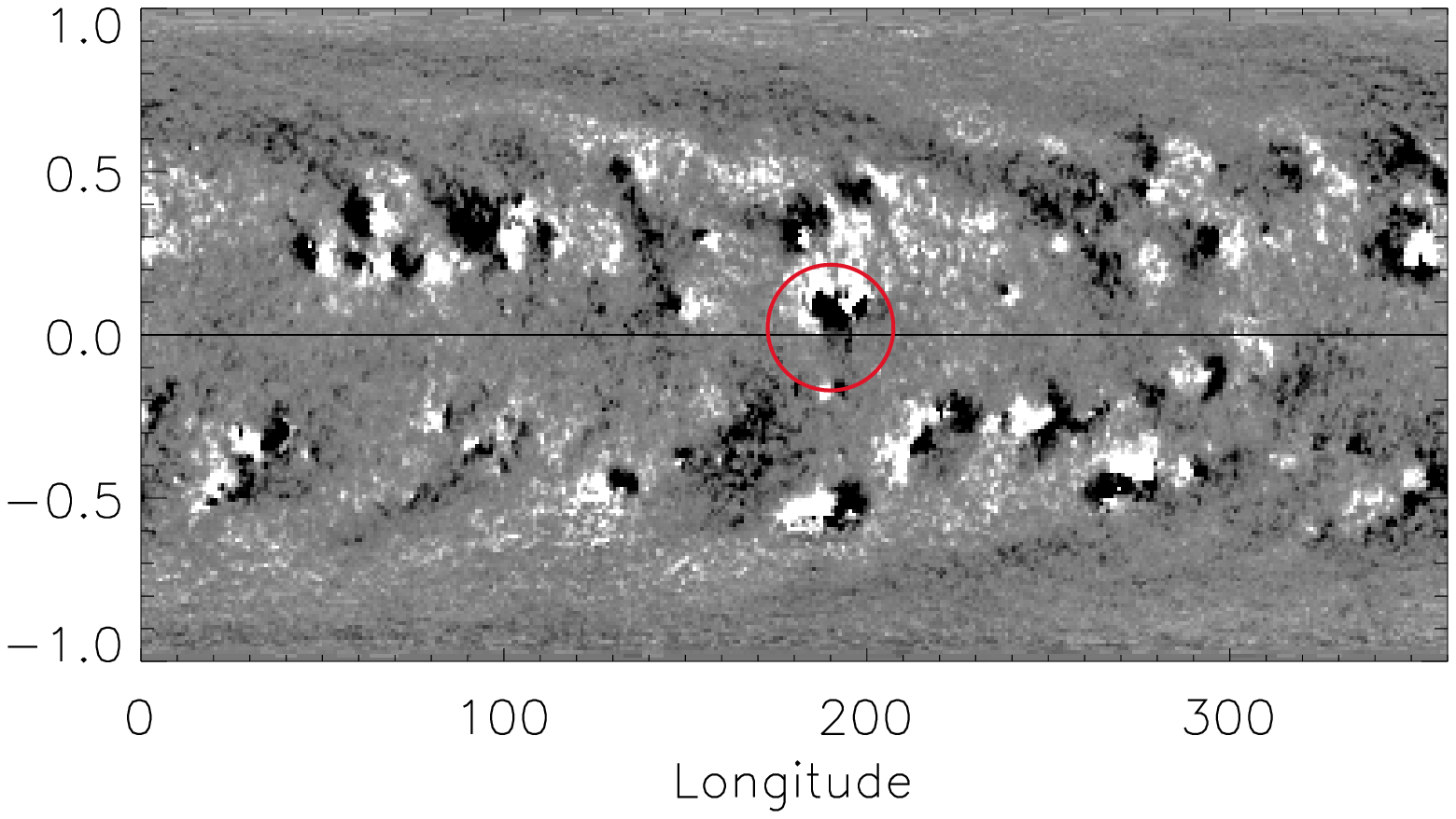}
\includegraphics[scale=0.45]{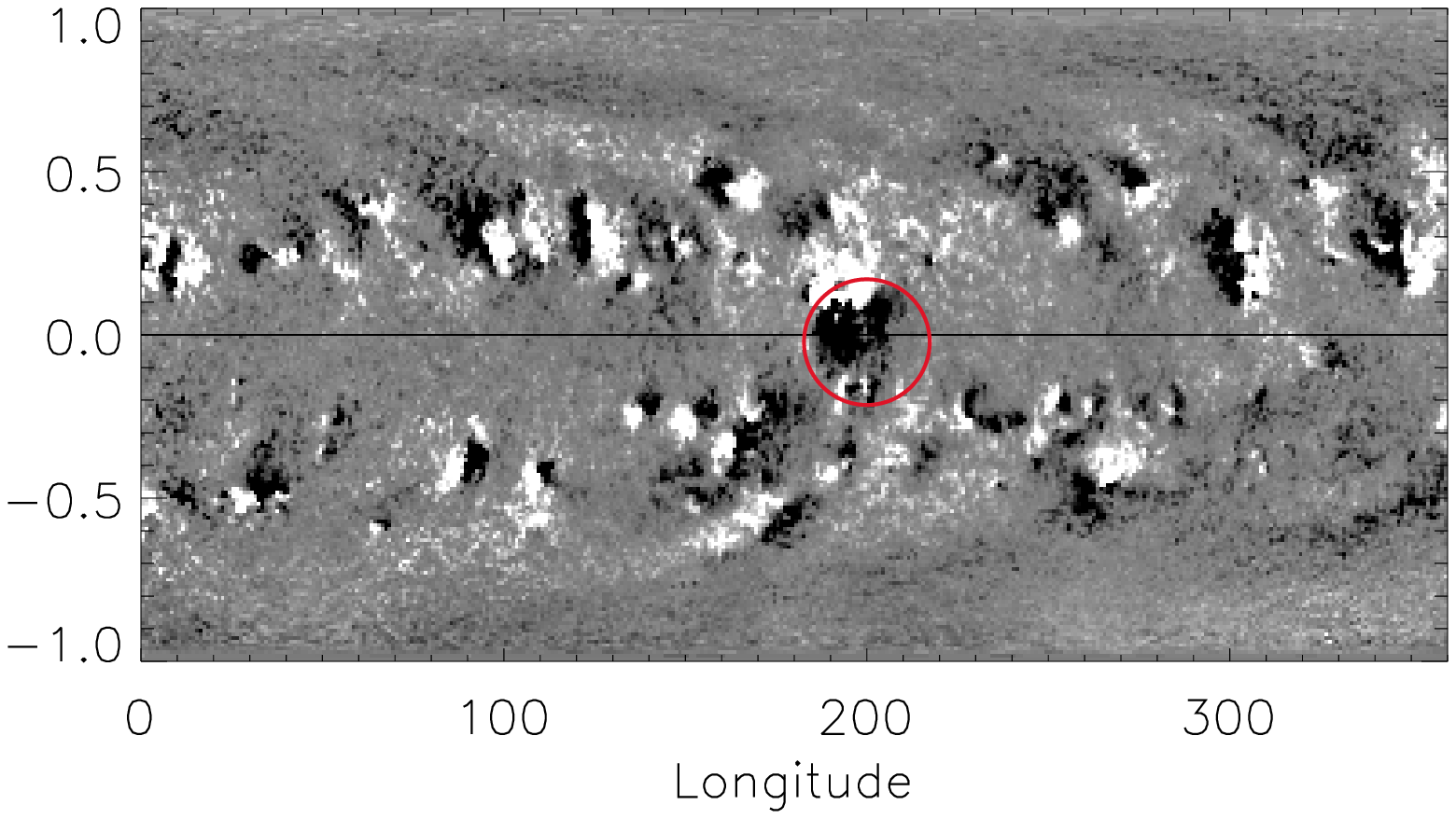}
\includegraphics[scale=0.45]{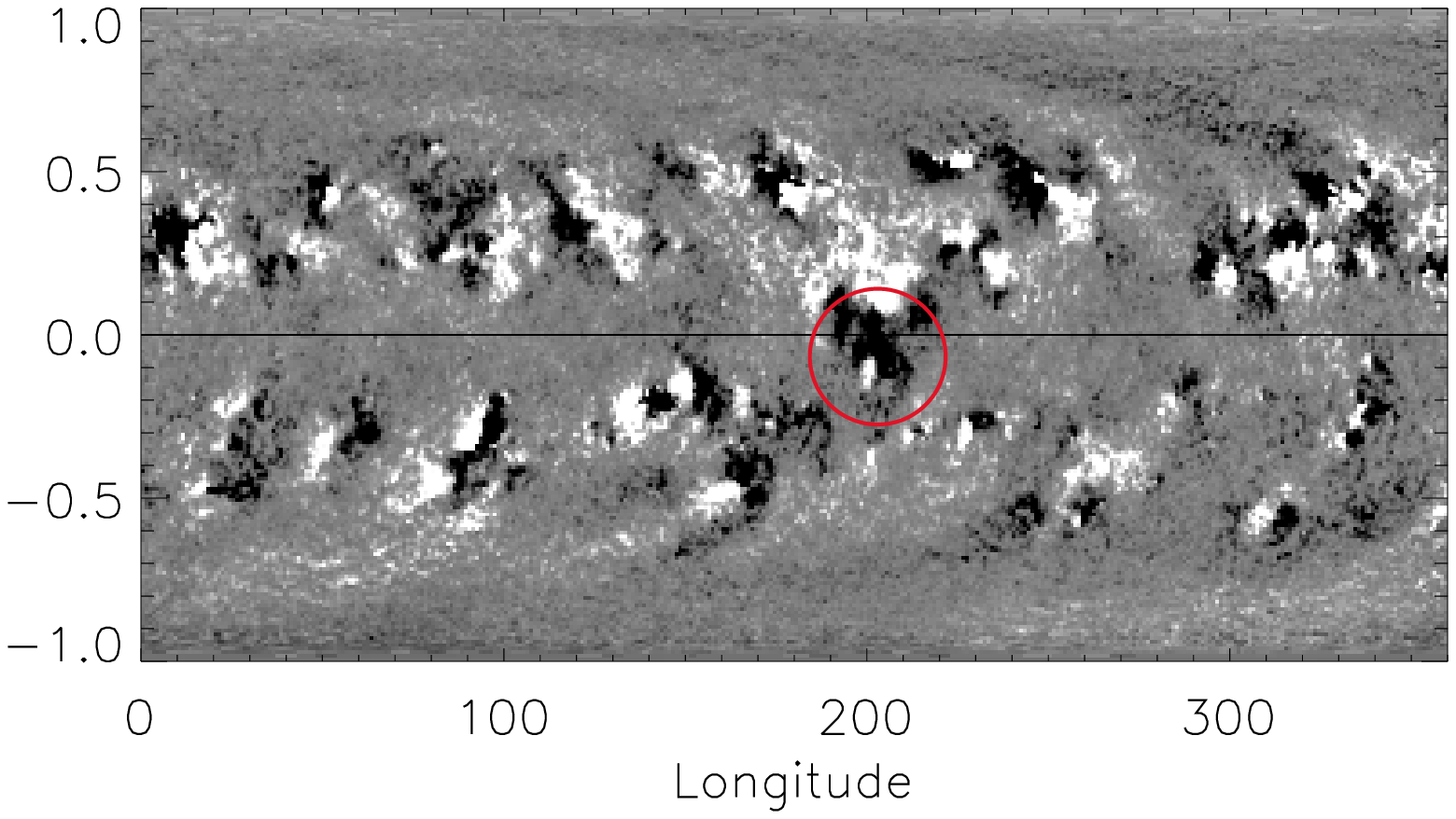}
\includegraphics[scale=0.45]{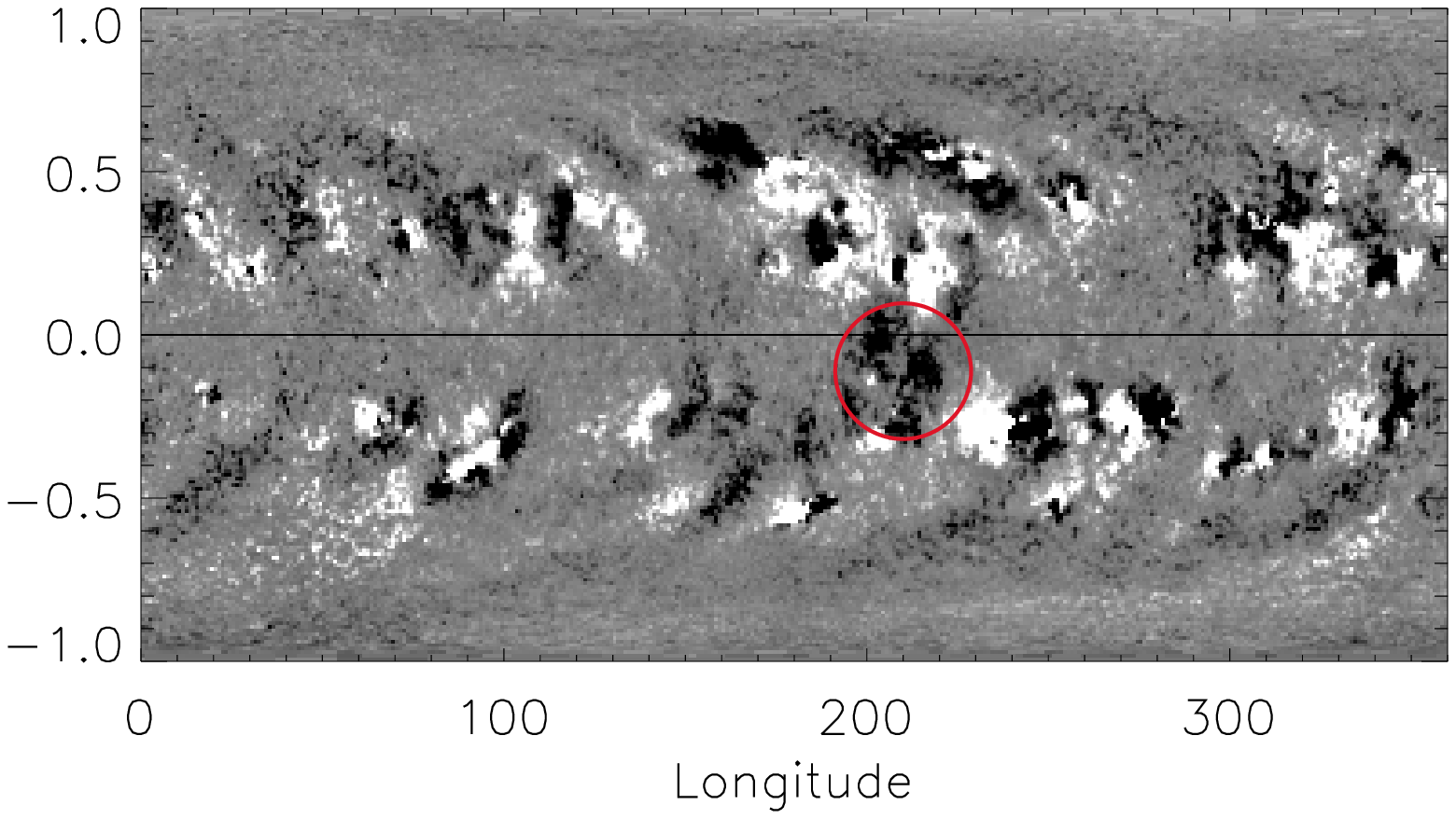}
\caption{Kitt Peak synoptic magnetograms for Carrington rotations 
1684-1688 (starting dates: 1979, July 17, August 13, September 9, 
October 6, November 3) are shown. The red ellipse outlines an 
example of non-diffusive cross-equatorial transport of magnetic flux. 
In this case the negative flux of a highly tilted bipolar region 
emerges across the equator. }
\label{fig:CR1}
\end{center}
\end{figure}

\begin{figure}[h!]
\begin{center}
\includegraphics[scale=0.45]{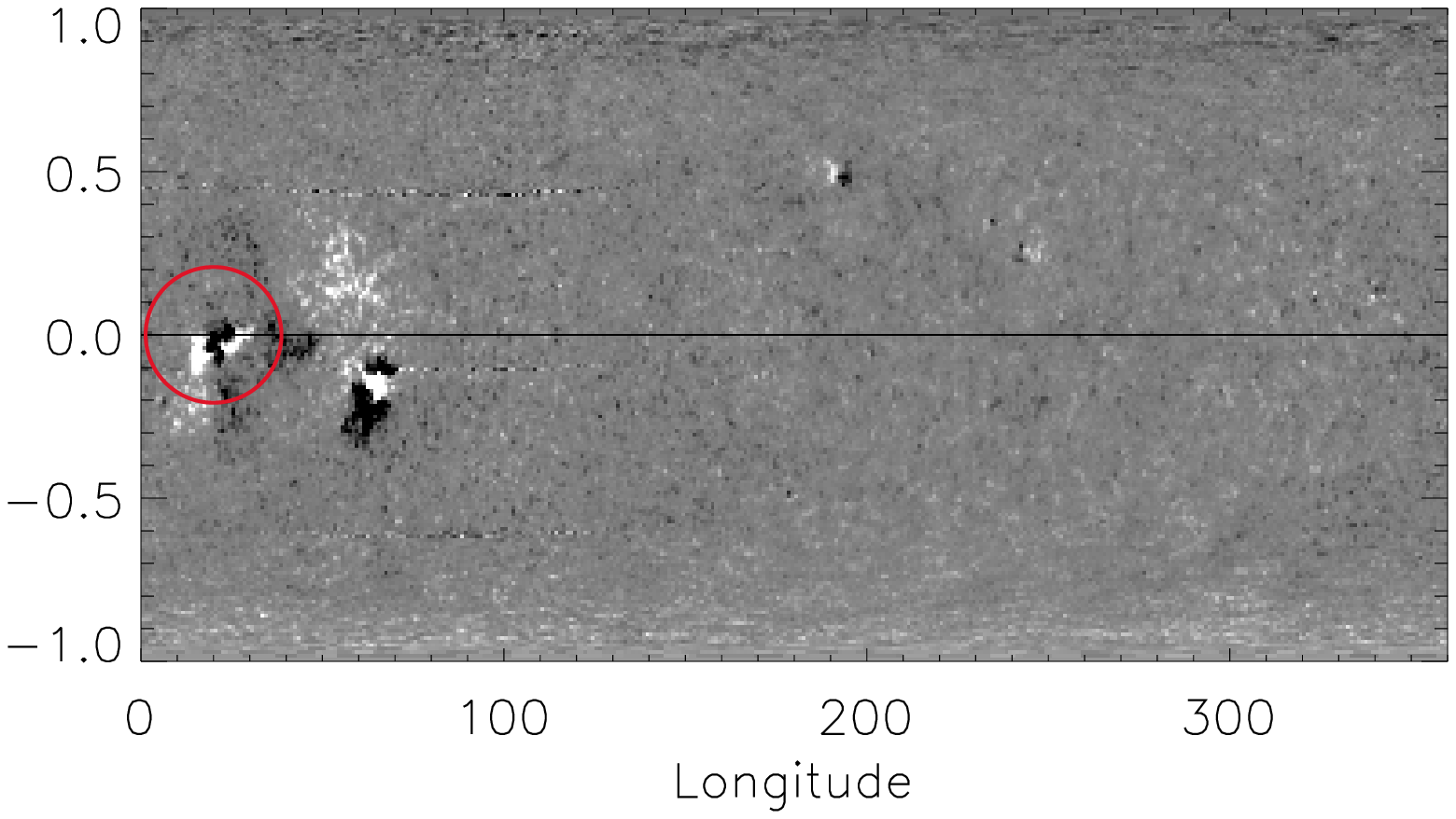}
\includegraphics[scale=0.45]{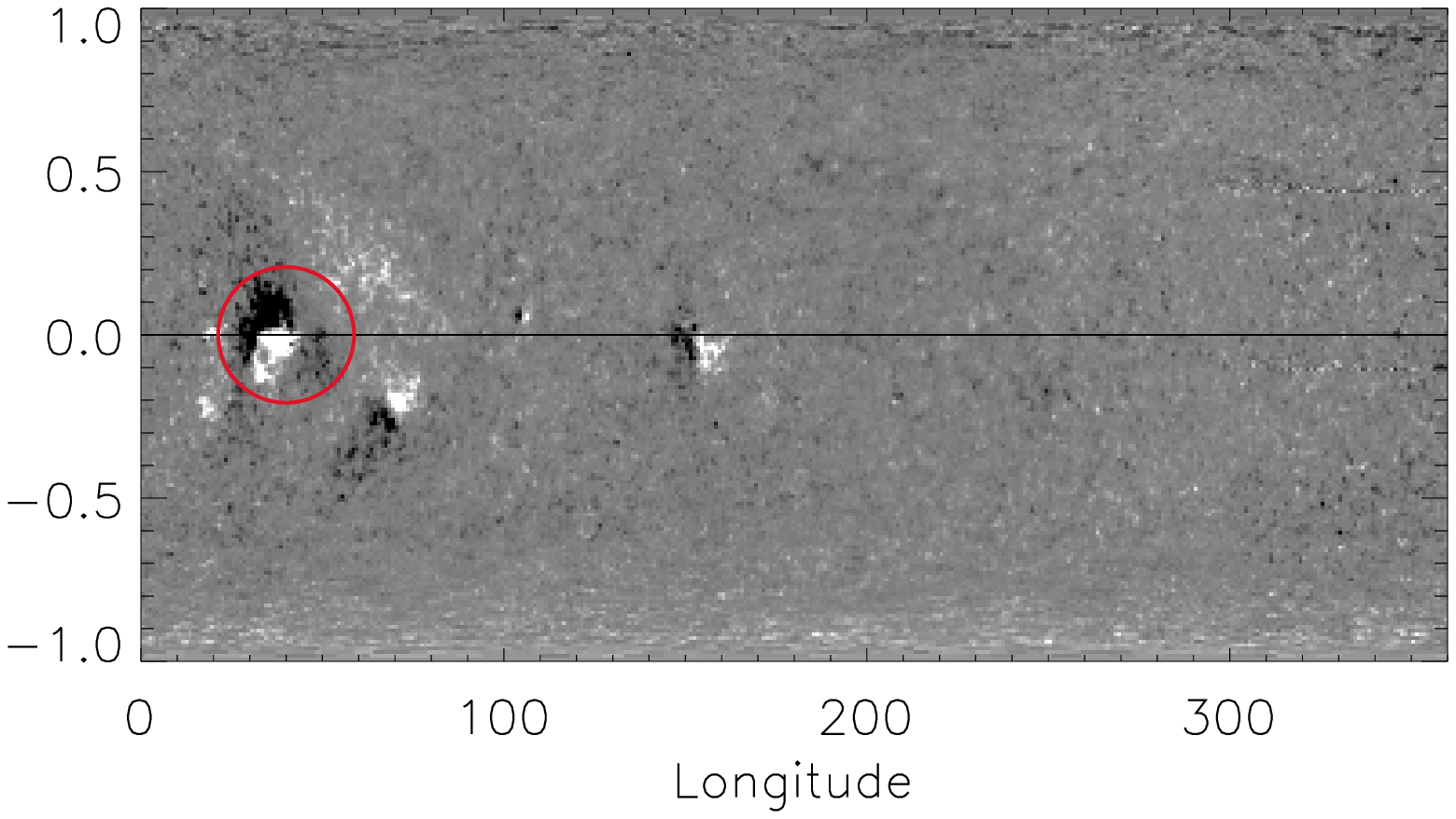}
\includegraphics[scale=0.45]{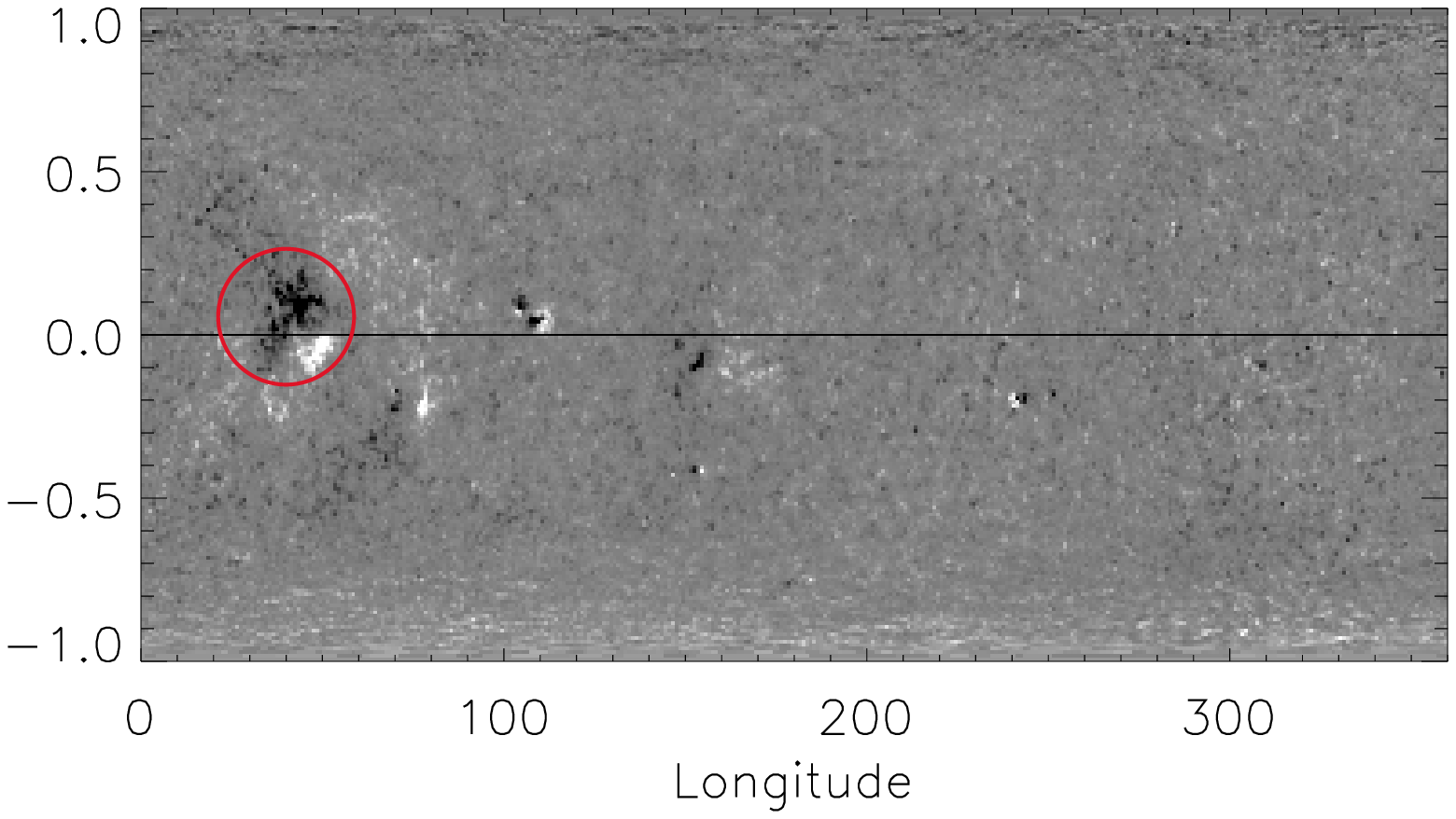}
\includegraphics[scale=0.45]{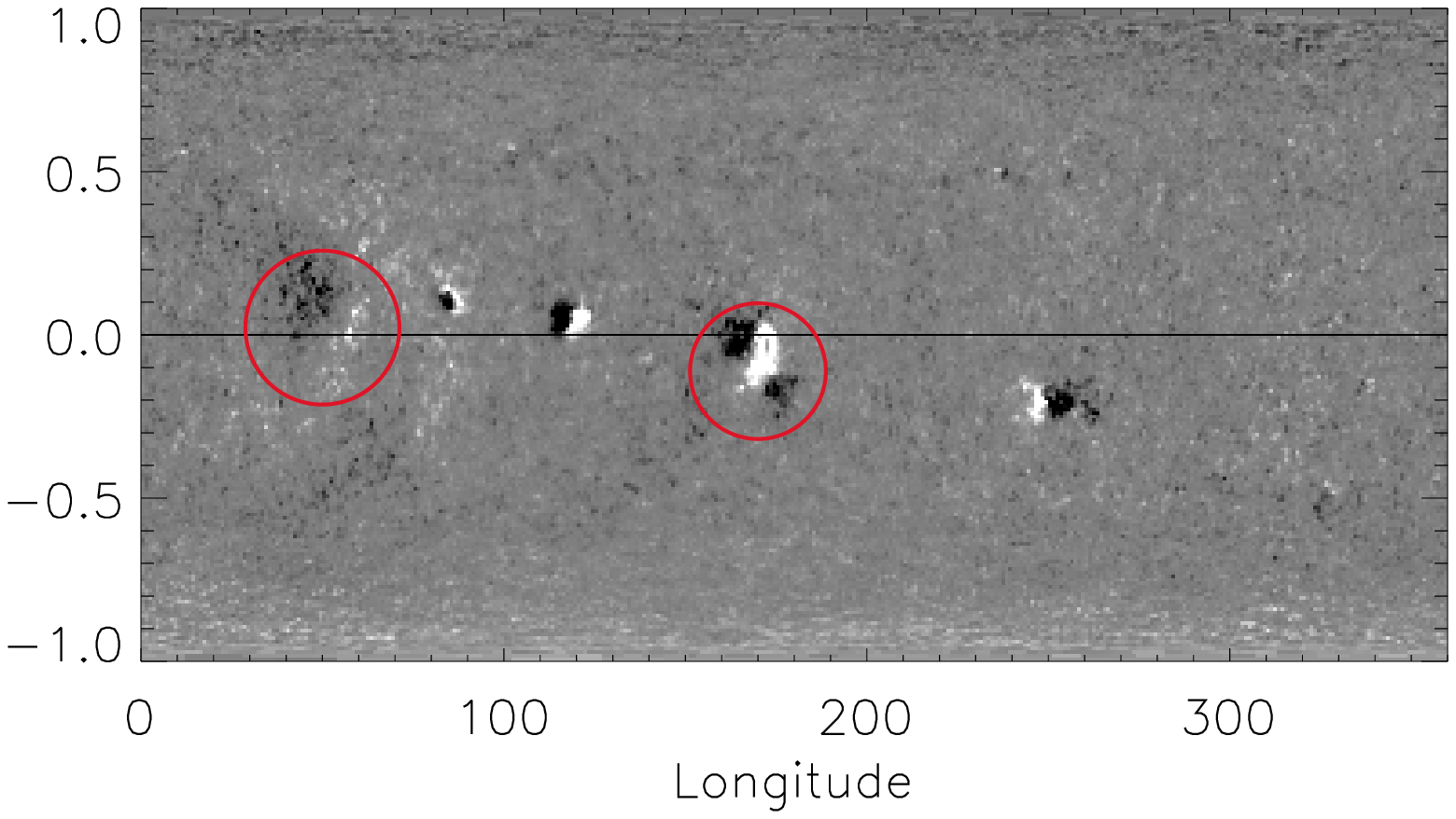}
\includegraphics[scale=0.45]{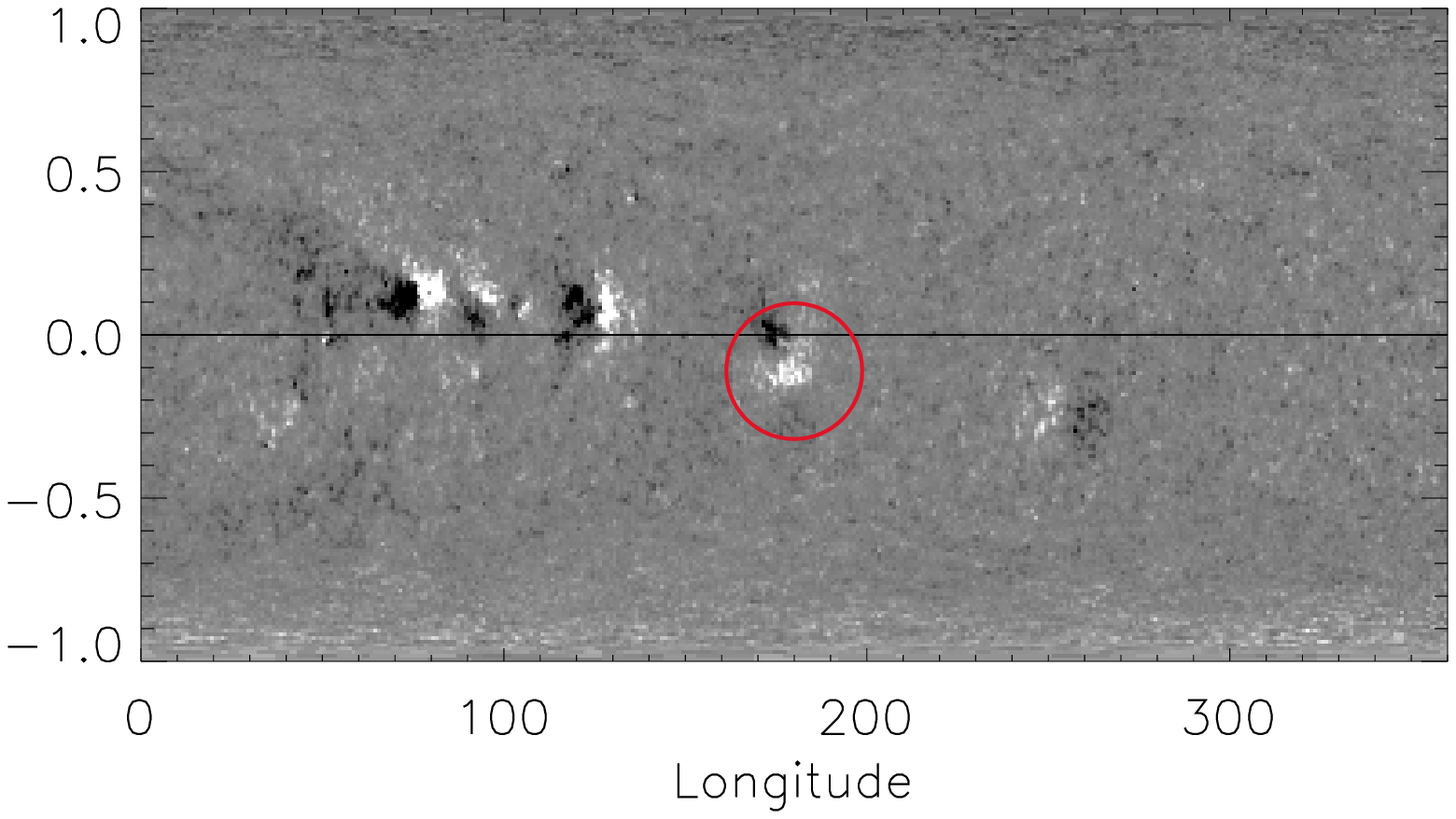}
\caption{Kitt Peak synoptic magnetograms for Carrington rotations 
1771-1775 (starting dates: 1986, January 14, February 10,  March 9, April 6, May 3).
The red circle outlines a magnetic bipole which emerges near the 
equator. Because it is highly tilted, the positive magnetic flux is almost entirely in the 
southern hemisphere and the negative flux is in the northern hemisphere. A similar, but weaker, 
event with a smaller tilt angle occurs towards the end of this event.}
\label{fig:CR2}
\end{center}
\end{figure}

Figure~\ref{fig:CR1} shows the evolution of the magnetic field which produces the cross-equatorial 
flux plume in 1980 circled in Figure~\ref{fig:mbf}. A large bipolar region has emerged with 
the positive polarity flux in the northern hemisphere and the negative
polarity extending to both sides of the equator. The axis of the bipolar pair,
the line connecting the centers of the opposite polarities, is inclined at almost 
90 degrees to the equator. In this particular case the north-south orientation of 
the bipolar pair is opposite to that given by Joy's law for this cycle.
A second event that occurred in 1986,
in the declining stages of cycle 21, is shown in Figure~\ref{fig:CR2}. Again we have flux 
emergence across the equator; however in this case the north-south alignment is in accordance 
with Joy's law. The latitudinal alignment of the bipolar groups is important because it determines
whether negative or positive flux is transported into the northern hemisphere. For cycle 21, 
transporting positive flux into the northern hemisphere acts to weaken the net flux in each 
hemisphere at the subsequent minimum, whereas transporting negative flux enhances the net flux
at minimum. The two events therefore mostly cancel each other for this cycle, 
as will be discussed in Section 4. 
There are several weak cross-equatorial flux plumes around 1990, before another prominent event 
occured in 2002.

\subsection{Emergence near the equator}
\begin{figure}[h!]
\begin{center}
\includegraphics[scale=0.45]{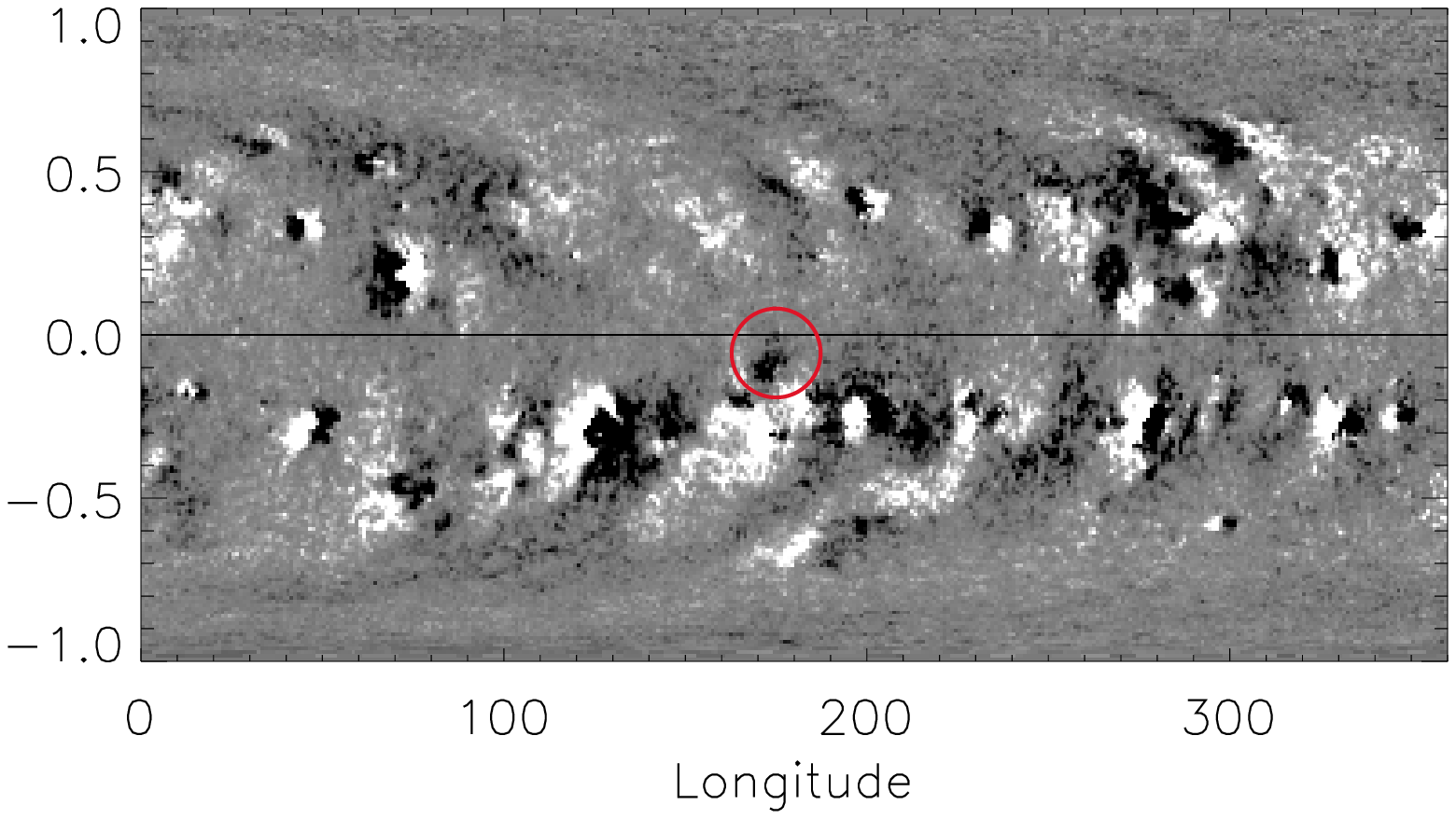}
\includegraphics[scale=0.45]{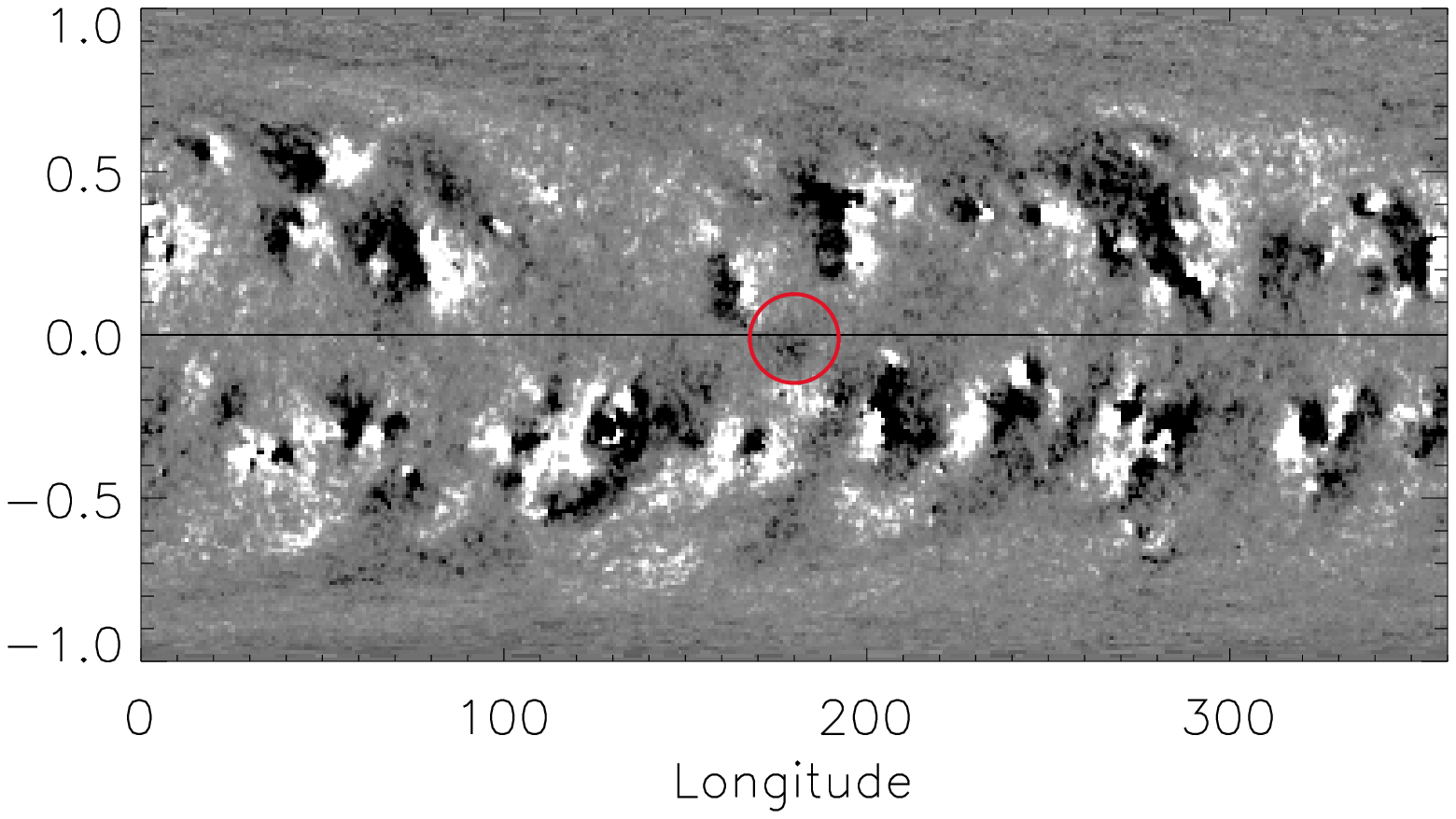}
\includegraphics[scale=0.45]{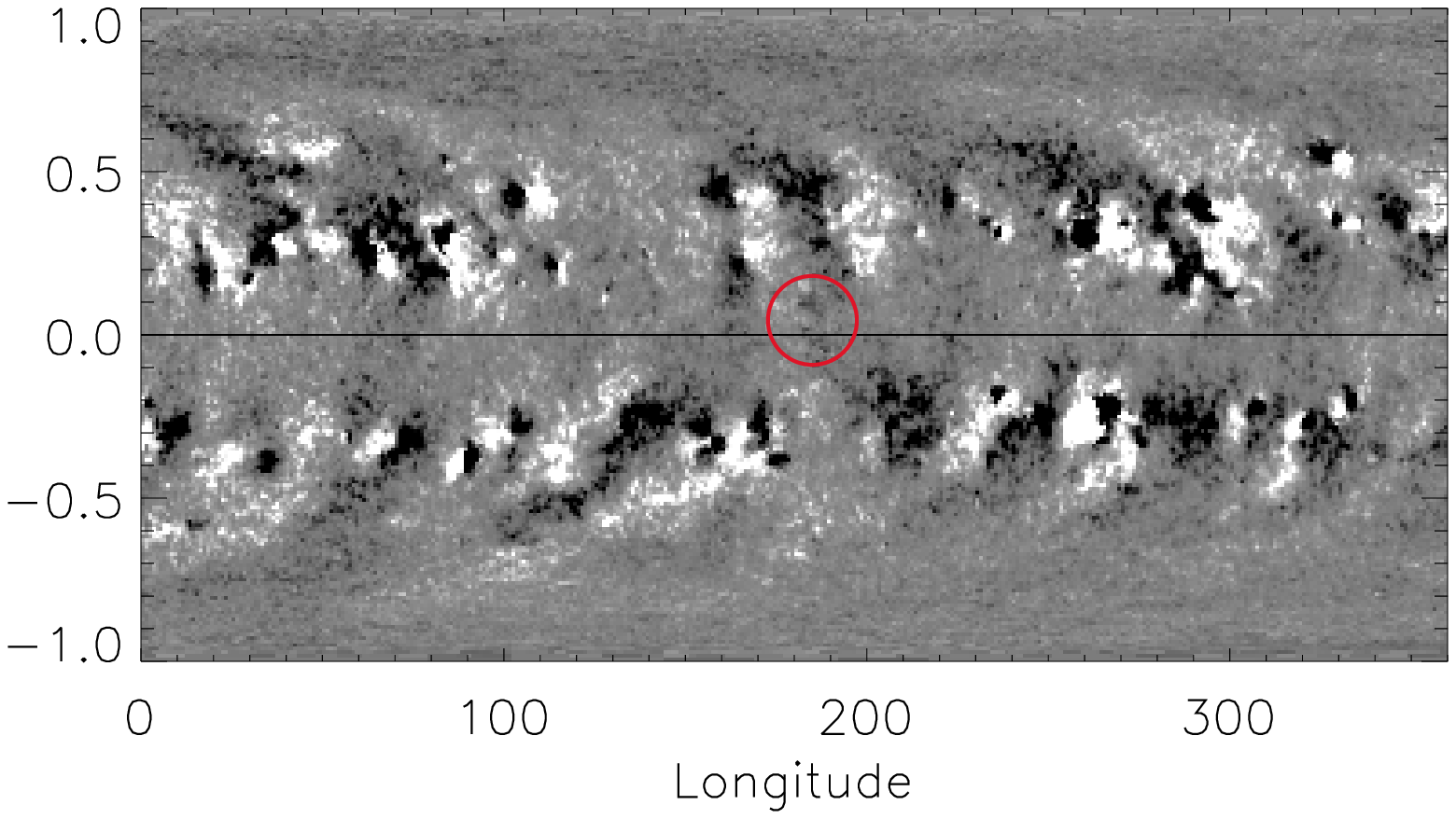}
\caption{Kitt Peak synoptic magnetograms for Carrington rotations 
1960-1962 (starting dates: 2000, February 25, March 23, April 19).
The red circle outlines negative flux which has emerged near the equator
as part of a highly tilted bipolar group. In the course of its evolution,  
the negative flux is transported across the equator.}
\label{fig:CR3}
\end{center}
\end{figure}

A second type of event is shown in Figure~\ref{fig:CR3}. Here  
a weaker, but still highly tilted, bipolar region emerges close to the 
equator. Both polarities emerge in the 
southern hemisphere, with the negative flux being closer to the equator.
Negative flux is transported across the equator, presumably 
driven by cross-equatorial flows. This leads to a substantial amount of 
flux crossing the equator.

\section{Measuring the cross-equatorial fluxes}
The calculation of the cross-equatorial flux transport of magnetic flux
is discussed by \cite{Durrant04}, who also estimated the diffusive component of the 
cross-equatorial flux transport during cycle 22. 
The net (signed) magnetic flux in the northern hemisphere is given by 
\begin{equation}
F^{\mathrm{NH}}= \int_{\mathrm{NH}} B_r dA,
\end{equation}
and in the southern hemisphere by 
\begin{equation}
F^{\mathrm{SH}}= \int_{\mathrm{SH}} B_r dA.
\end{equation}
Because $\nabla \cdot {\bf{B}}=0$ these must satisfy 
$F^{\mathrm{NH}}=-F^{\mathrm{SH}}$.
To reduce the noise we define $F=(F^{NH}-F^{SH})/2$, i.e. 
$F$ is an estimate of the flux in the northern hemisphere based on observations 
over both hemispheres.
The net magnetic flux transport across the equator at the solar surface is then 
determined by ${d F}/{dt}$.

\begin{figure*}[H]
\begin{center}
\includegraphics[scale=0.4]{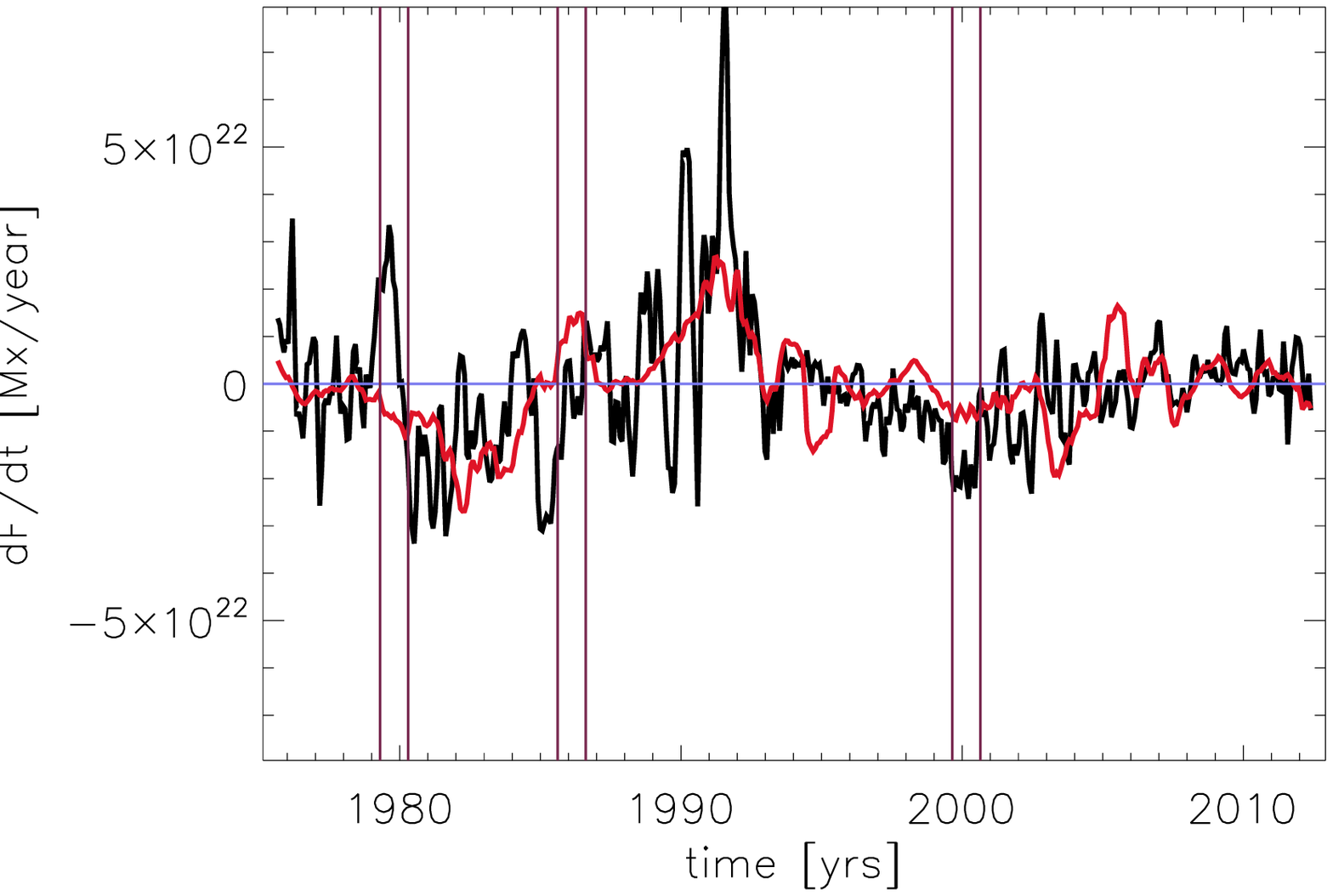}\includegraphics[scale=0.4]{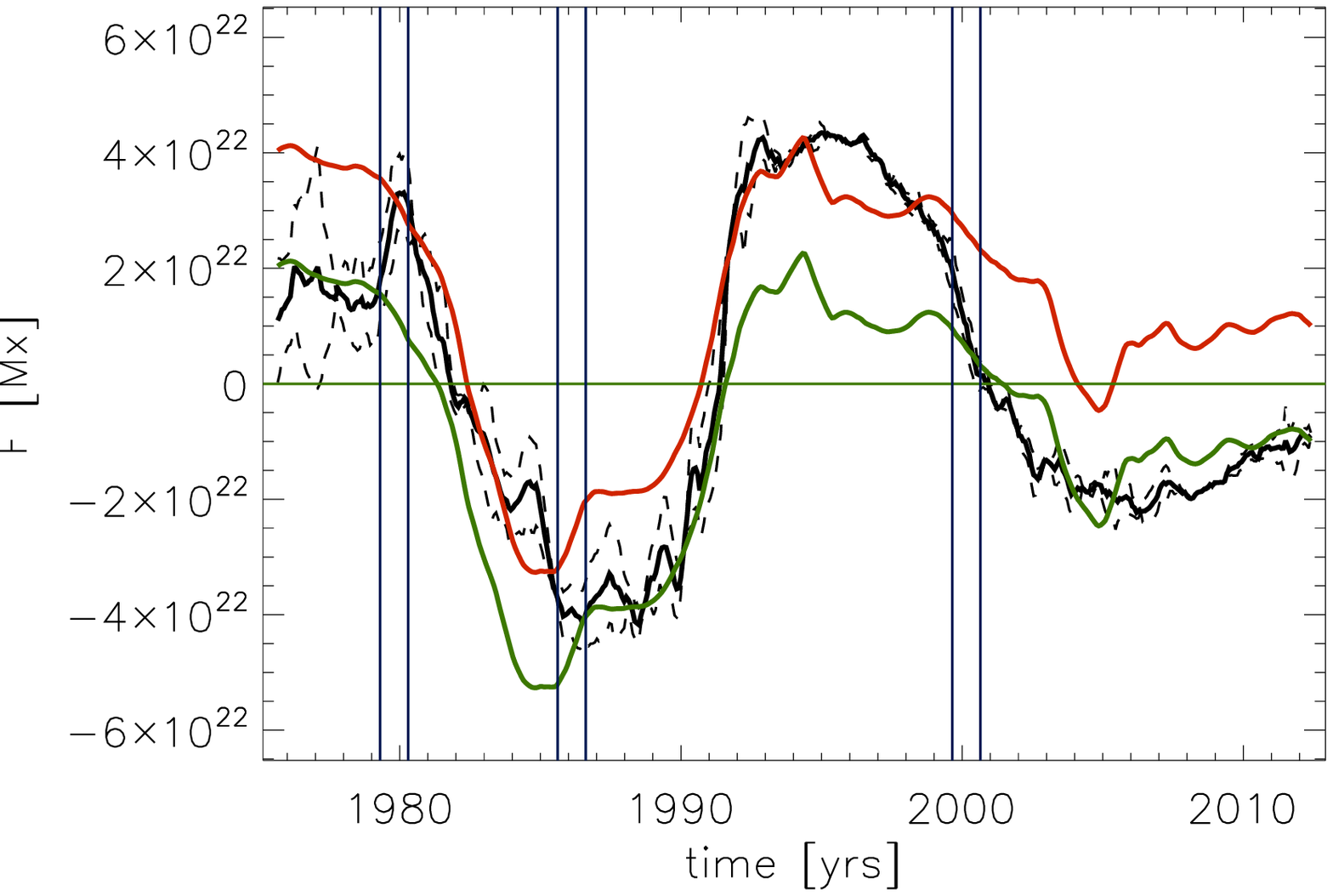}\\
\includegraphics[scale=0.4]{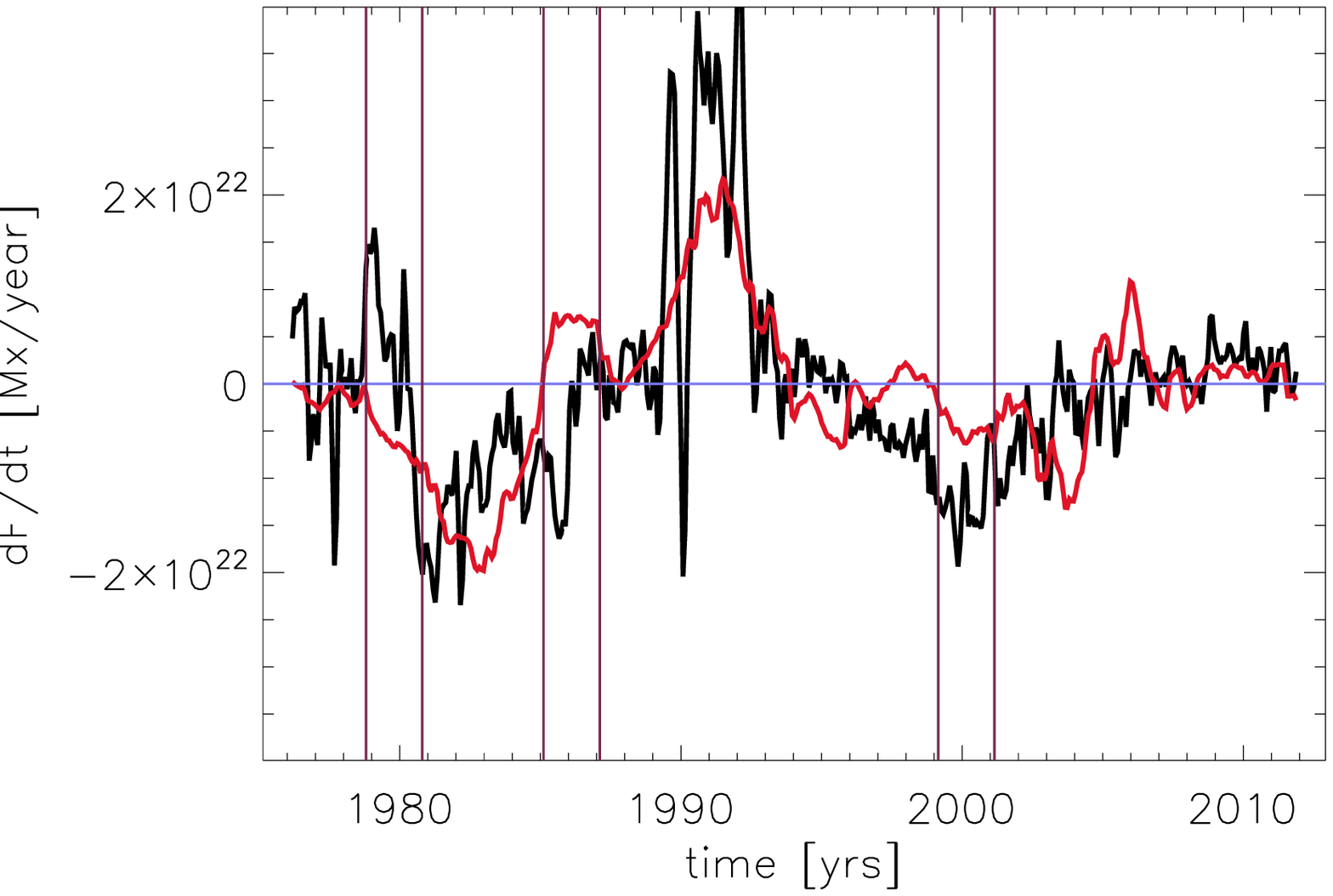}\includegraphics[scale=0.4]{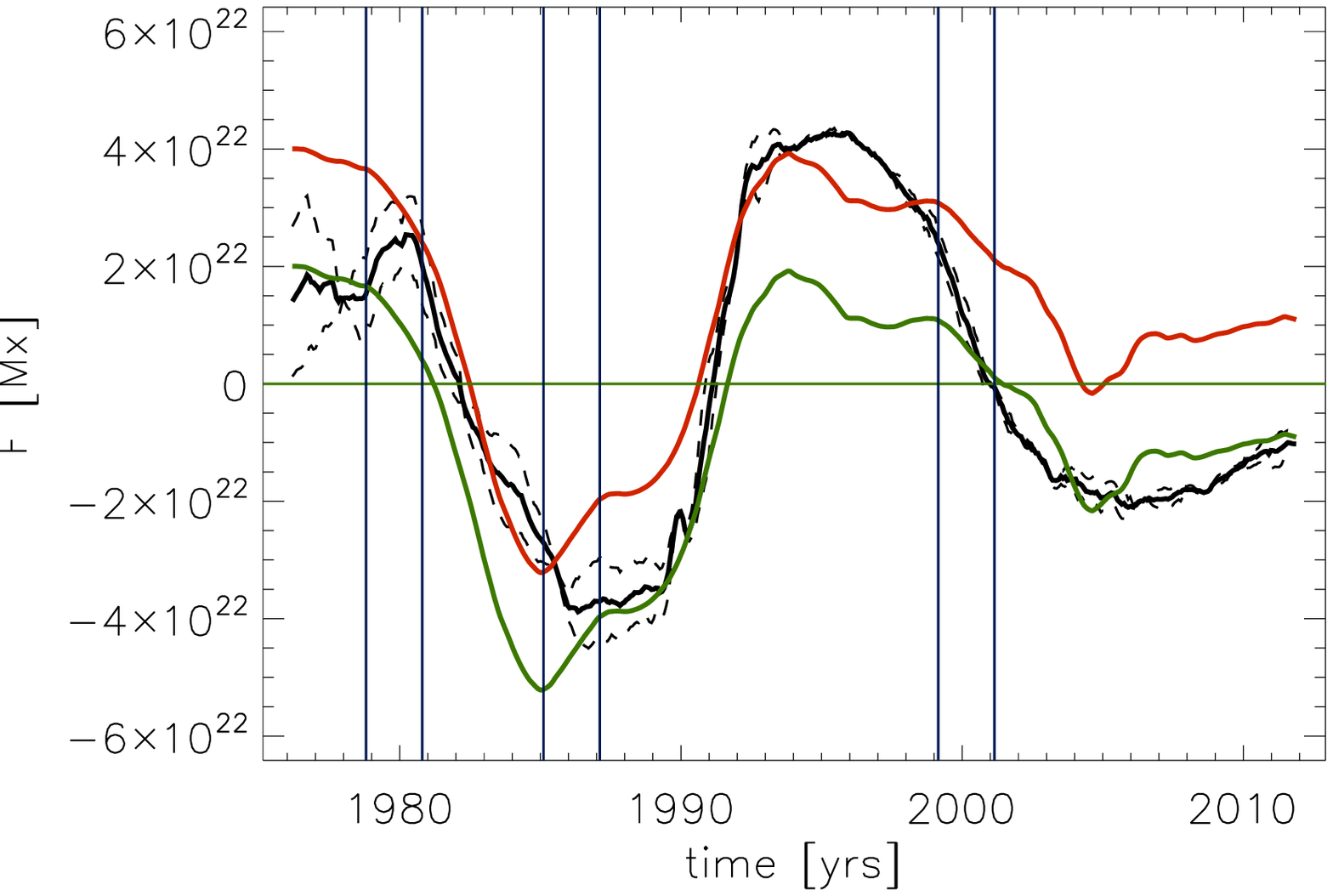}\\
\includegraphics[scale=0.4]{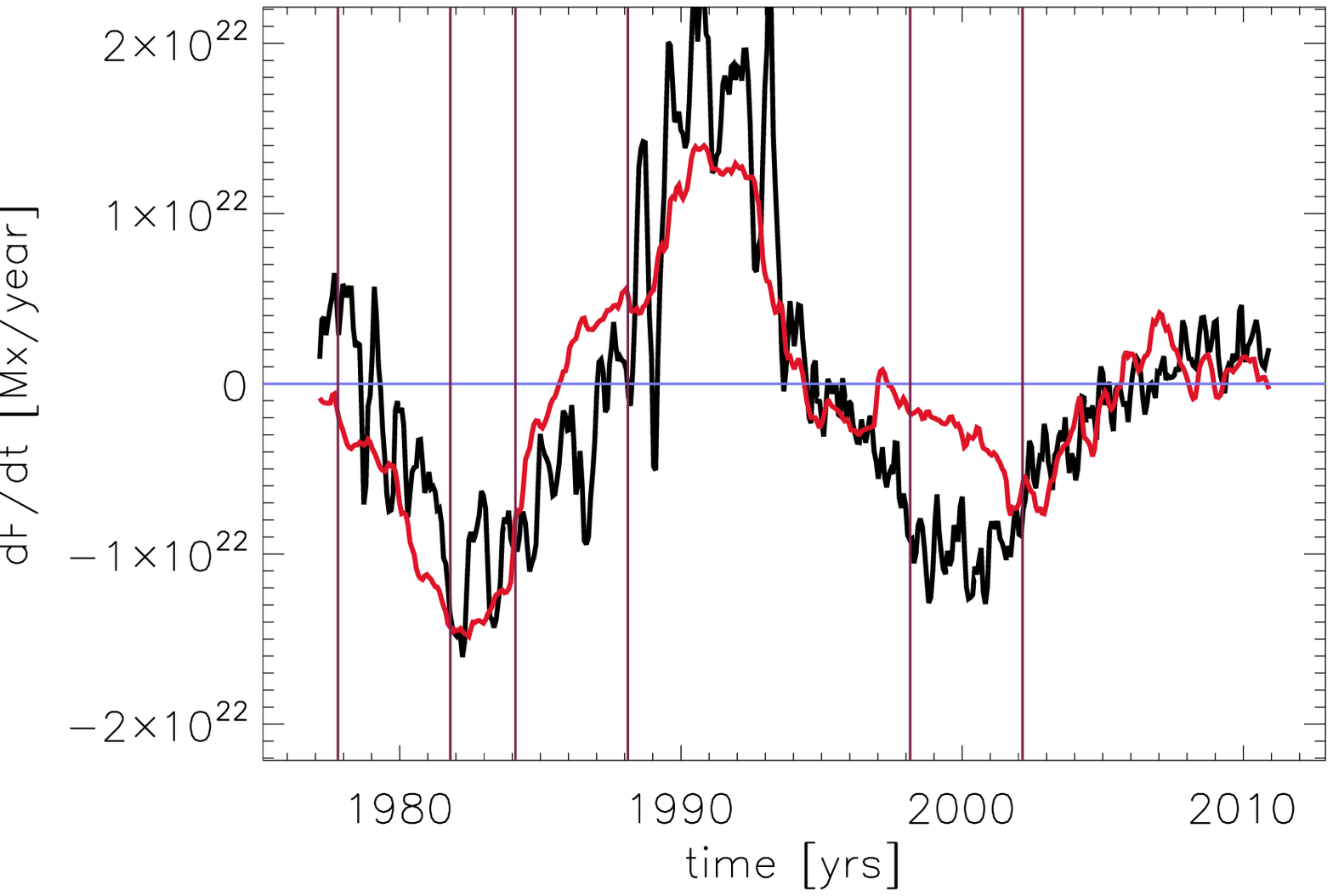}\includegraphics[scale=0.4]{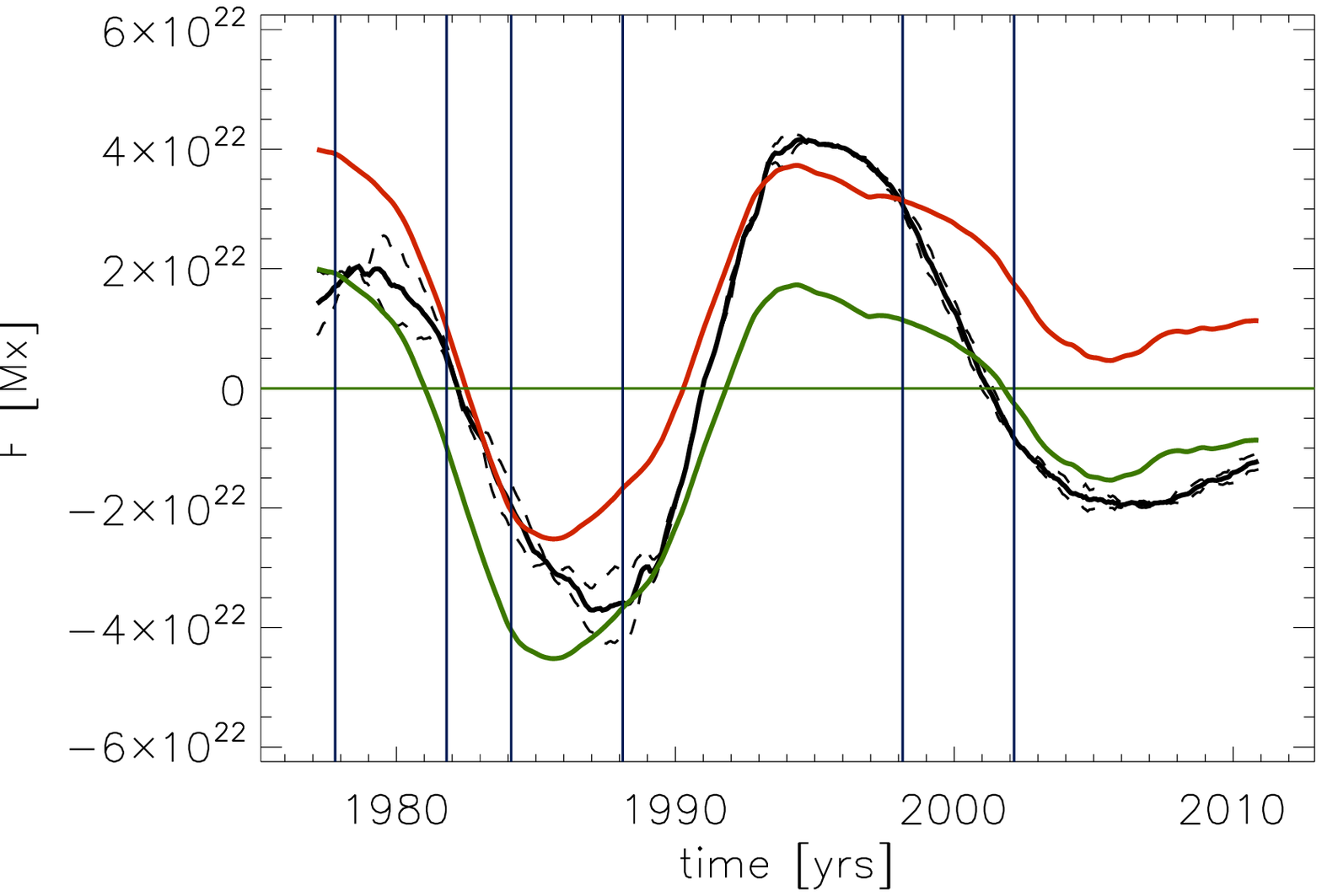}
\caption{Net cross equatorial transport of magnetic flux as a function
of time (left panels). The black curves show the total net flux transport, ${dF}/{dt}$ 
averaged over 1 year (top panel), 2 years (middle panel), and 4
years (bottom panel).  The red curves show the 
diffusive component of the flux transport  
$\frac{dF}{dt}_{\mathrm{diffusive}}$, see Equation~3.
The right panels show $F^{\mathrm{NH}}$, the magnetic flux in the northern hemisphere, 
and $-F^{\mathrm{SH}}$, the negative of the magnetic flux in the southern hemisphere (dashed curves), 
as well as their mean (solid black curve). For comparison the time integral of the diffusive flux is 
shown in red (assuming a starting value of $4\times 10^{22}$~Mx in 1976) and green
(assuming a starting value of $2\times 10^{22}$~Mx). 
The intervals outlined by the vertical lines indicate the periods affected by the three 
events shown in Figures~\ref{fig:CR1},~\ref{fig:CR2}, and~\ref{fig:CR3}, respectively.}
\label{fig:match}
\end{center}
\end{figure*}

Because we are numerically evaluating the time derivative, the measured 
cross-equatorial transport is noisy.
One obvious source of noise is the yearly apparent modulation of the 
polar fields in  Figure~\ref{fig:mbf}, which is an artifact 
introduced by the variation of the solar B-angle due to the inclination 
of the solar rotation axis to the ecliptic.
By averaging over 13, 27, or 54 Carrington rotations (approximately 
1, 2, and 4 years, respectively) this source of noise is substantially reduced.
The black lines in the left-hand panels of Figure~\ref{fig:match} 
show the time history of ${d F}/{d t}$.

We also estimate the amount of cross-equatorial flux transport which is due to
diffusion of magnetic flux across the equator. 
We consider the turbulent diffusion describing the random walk of magnetic 
features associated with supergranulation, averaging the magnetic field over 
supergranular scales using a box-car filter with a width of 24 Mm. 
We then calculate the latitudinal gradient of the magnetic field at the equator
using centered finite differences and estimate the diffusive cross-equatorial
flux transport as
\begin{equation}
\left(\frac{dF}{dt}\right)_{\mathrm{diffusive}}=-2 \pi \eta_{\mathrm{turb}}
             \frac{\partial B_r}{\partial \lambda}\mid_{\lambda=0}, 
\end{equation}
where $\lambda$ is latitude
and we assume $\eta_{\mathrm{turb}}=250$~km$^2$s$^{-1}$ \citep[][]{Cameron10}.
This diffusive component of the cross equatorial flux transport is  
shown by the red lines in the left-hand panels of Figure~\ref{fig:match}.
Explicitly, it is the amount of flux carried across the
equator by diffusion: how the flux arrives near the equator (before being 
transported across the equator by diffusion) will in general 
involve a mixture of emergence, advection and diffusion.

\section{Relation of polar flux and net flux in each hemisphere at minima}
\begin{figure}
\begin{center}
\includegraphics[scale=0.45]{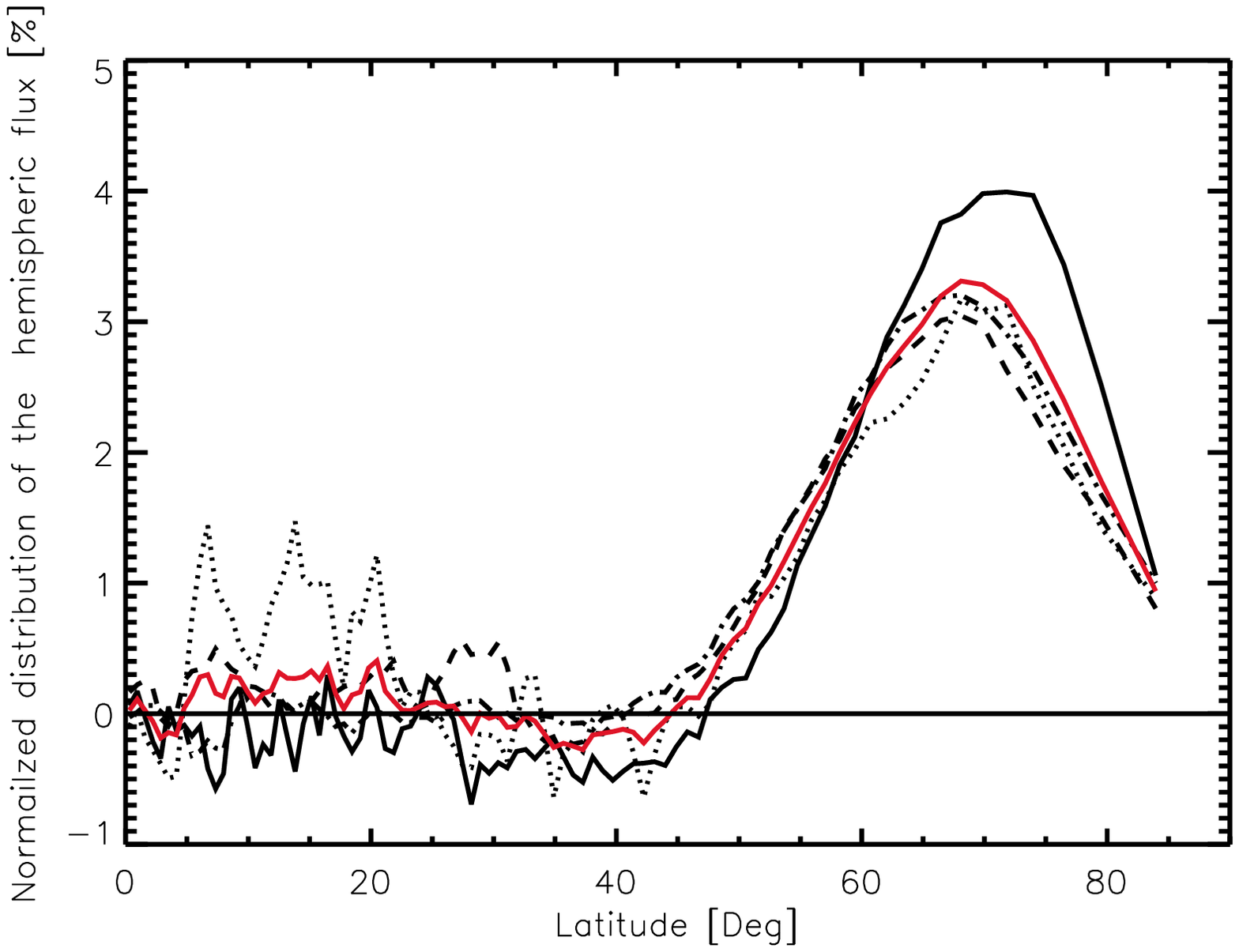}
\caption{Normalized distribution of the net flux in each hemisphere per degree as a function of latitude
for the minimum between cycles 23 to 24 (solid black),  cycle 22 to 23 (dash--dot),
21 to 22 (dashed) and 20 to 21 (dotted). In each case the fluxes have been averaged over 2 years 
centred on the minimum. The red curve shows the average over all the cycles.} 
\label{fig:polar}
\end{center}
\end{figure}
So far our discussion has concentrated on the net flux in each hemisphere. The more
common precursors used to make predictions of solar activity are the polar fields and 
open heliospheric flux at solar minimum. Figure~\ref{fig:polar} shows the contribution of each latitude 
to the net flux in each hemisphere for the last four minima (those for which observations 
are available).  At solar minimum most of the net flux in each hemisphere is
concentrated near the poles: on average over 78\% of the flux in each hemisphere is above 
60$^{\circ}$ latitude. The variation in the net flux in each hemisphere is thus a major factor
in determining the strength of the polar fields at minimum. This is because the by the time
that the activity has reached its minimum, the magnetic field 
distribution has had sufficient time to move toward an equilibrium for which the transport due to 
advection by
the meridional flow towards the poles is balanced by the diffusion of field away from the pole.
The relative contributions over the last four cycles do show some variability, but this is 
substantially less than the variability of the cycle strengths. Hence most of the cycle-to-cycle
variability in the polar field strengths is coming from changes in the net flux in each hemispere.
We similarly expect the open flux at minimum to be highly correlated with the net flux 
in each hemisphere.

\section{Discussion}
Comparison of the black and red curves on the 
left panels of Figure~\ref{fig:match} show that most of the net flux transport
across the equator is represented well by diffusion with  $\eta_{\mathrm{turb}}=250$~km$^2$s$^{-1}$.
This value is in accordance with observations \citep[see Table 6.1][]{Schrijver00},
simulation \citep{Cameron11}, and also with what is required to reproduce the evolution of the Sun's
large-scale magnetic field using the surface flux transport model with a meridional
flow profile consistent with observations \citep{Cameron10}. 
While the curves 
are similar, there are also differences which represent 
the non-diffusive component of cross-equatorial flux transport. The  
examples of cross-equatorial flux plumes discussed in Sections 2.2 and 2.3 are clearly
associated with such non-diffusive transport. Figure~\ref{fig:match} also
indicates that there are more cross-equatorial flux plumes than the examples
discussed, yet these events are among the most prominent. We note that we
chose some of the examples based upon their signature in Figure~\ref{fig:match},
although they only have a weak signature in Figure~\ref{fig:mbf}. We also comment that 
some apparently strong cross-equatorial flux plumes in the magnetic butterfly diagram,
such as the event in 2002, show only a weak signature in 
Figure~\ref{fig:match}. In this case, the positive polarity flux plume is preceded
by a weaker but longer lived negative polarity flux plume. The time averaging
then smoothes these two, erasing the signature of both events.

The impact of the cross-equatorial flux plumes can be seen clearly in the evolution 
of the net flux in the northern hemisphere,
$F$, shown in the right panels of Figure~\ref{fig:match}. 
The amount of flux in the northern hemisphere that,
     at any given time, which was carried across the equator 
     by diffusion alone
can then be calculated as $\int ({d F}/{d t})_{\mathrm{diffusive}} dt+F_0$.
The value of $F_0$ is a time-independent offset to the net flux in each hemisphere.
If the transport were only diffusive, then a single offset (the amount of flux in the 
northern hemisphere at the start of the time series $F_0=2\times 10^{22}$~Mx)
would make the curve for the diffusive transport, (shown by the green curve) 
match the total flux transport (black curve)  throughout the entire period. Instead it is only 
consistent with the total transport until about 1980. 
At this time there is substantial cross-equatorial transport due to the event shown in 
Figure~\ref{fig:CR1}. The net flux in the northern hemisphere, which has not
yet reversed, is strengthened.  This transport is non-diffusive, and so the 
evolution of the net flux in the northern hemisphere no longer follows the
the purely diffusive green curve.  The evolution after 1980  is again mostly diffusive --
the field evolves along the red curve in  Figure~\ref{fig:match}  which gives the
evolution from $F_0=4\times 10^{22}$~Mx being the sum of the original $F_0=2\times 10^{22}$~Mx
and the $2\times 10^{22}$~Mx injected from the non-diffusive transport.
This red curve is inconsistent with the evolution before 1980, and has a mean which is 
higher than the observations. It does however match the observed evolution from 1980 until 
the event shown in Figure~\ref{fig:CR2} occurs in 1986 and the net flux in the northern hemisphere drops 
(non-diffusively) by about  $2\times 10^{22}$~Mx (back to the green curve).
In this particular case, the two events thus cancel each other and the net result is 
similar to purely diffusive transport. Thereafter, the evolution is again mainly due to diffusion until 1990. 
There is then a period when a number of weak cross-equatorial positive magnetic plumes can be 
seen crossing the equator in the magnetic butterfly diagram. Although these 
plumes individually do not carry much flux, their combined effect is substantial
and the evolution switches again to follow the red curve. From 1992 to
2000 the evolution is roughly in accordance with purely diffusive transport.
In 2000 another event transports flux across the equator and the evolution
switches back to mostly diffusive transport at about the level of the green curve  
with $F_0=2\times 10^{22}$~Mx.

Each of the cross-equatorial transport events studied here transports 
about $2\times 10^{22}$~Mx of flux across the equator (presumably due to chance and 
because we focussed on the largest events).
Consistent with the synoptic magnetograms, this is a significant fraction 
of the total unsigned flux associated with a large active region 
\citep[][ Table 5.1, gives $1 \times 10^{22}$ to $6\times 10^{22}$~Mx as
the range for the unsigned flux of a large active region]{Schrijver00}. 
The diffusive flux integrated over a cycle can be estimated from Figure~\ref{fig:match} 
to be about $6 \times 10^{22}$~Mx, which is only about 3 times larger than the 
flux carried by the largest of the observed cross-equatorial flux plumes.
It follows that a single cross-equatorial flux plume introduces
a change in the total flux transported across the equator during a 
cycle of approximately $\pm 30$\%. The effect on the polar fields is about twice this, 
that is a change of $\pm 60$\%. This
is because the flux which crosses the equator first cancels the 
existing polar fields, and then reverses them.  The amount of flux required
to cancel the existing flux is independent of the total amount of flux which is 
transported across the equator. The flux which has been transported across the
equator only contributes to the reversed net hemispheric flux after the (fixed) amount of flux
which is required to cancel the old flux is surpassed.


In the cases studied here, we have events which weaken the net hemispheric flux 
at the subsequent minimum (the cross-equatorial flux plume in 1980) as well 
as those which strengthen it (the events in 1986 and 2000). 
We also have an example where the event occurred before maximum, as well as one
which occurred near (but before) the minimum.  Because the events
cancel in some cycles, the net flux carried by the plumes in the cycles studied here is about 
$\pm 10^{22}$~Mx/cycle. A failure to account for this flux in,
for example, surface flux transport models will cause the model to undergo a random walk
away from the observed flux. For modeling cycles 15 to 21, as for example in \cite{Cameron10},
where the tilt angles of individual bipolar groups is not available and a (cycle-dependent) Joy's
law is assumed, the RMS error introduced by not modeling the types of events is 9.8$\times 10^{21}$~Mx. 
This was calculated assuming a random walk of $\pm 10^{22}$~Mx/cycle and noting that the initial 
field strength of the model was a free parameter. This is to be compared 
with the range of fluxes, from 2.5$\times 10^{22}$~Mx to 6$\times 10^{22}$~Mx, for the open flux at
minima reported by \cite{Lockwood03}. Thus SFT simulations for a few solar cycles are justified, 
even if they neglect the cross-equatorial plumes studied here.
However the fact that the error is associated with a random walk means that the RMS 
error increases with the number of cycles simulated, and occassionally the errors introduced by
a single cycle will be important.
Because the number of large events is small, we are unable to comment further on their statistics.

\section{Conclusion}

The main result of this work is that a single cross-equatorial flux plumes can 
affect the net hemispheric flux of the following minima by up to 60\%. Furthermore,
whether the effect is to enhance or weaken the net flux depends
on whether the tilt is positive or negative. The number of cross-equatorial flux plumes
studied here is too small to evaluate whether there is a preference for 
constructive or destructive events.
Therefore, any prediction of the net flux in each hemisphere is subject to a large uncertainty.
Moreover, the strong correlation between the minima of the geomagnetic 
$aa$ indices (a proxy for the Sun's axial dipole moment) and the amplitude 
of the next cycle \citep{Wang09}, indicates that this uncertainty should
carry over directly into an uncertainty in predictions of the level of the 
Sun's activity \citep[e.g.][]{Schatten78, Jiang07, Yeates13}.  

Not every cycle may have such a large cross-equatorial flux plume, and some cycles
might have several smaller flux plumes which partially cancel each other. 
Nonetheless these events 
probably  will have a substantial random impact on the strength 
of the  net hemispheric flux. In rare instances it is possible that the flux 
transport across the equator could be effectively reduced by 50\% and 
lead to no net flux during the subsequent minimum. This might be  
a scenario for the origin of grand minima in the Babcock-Leighton model of the solar dynamo. 
In any case even single flux plumes, such as occurred in cycles 21, 22 and 23, can have a 
large impact on the evolution of the polar fields.

\begin{acknowledgements}
NSO/Kitt Peak data used here are produced cooperatively by NSF/NOAO, 
NASA/GSFC, and NOAA/SEL. This work utilizes SOLIS data obtained by the NSO Integrated Synoptic 
Program (NISP), managed by the National Solar Observatory, which is 
operated by the Association of Universities for Research in Astronomy (AURA), 
Inc. under a cooperative agreement with the National Science Foundation.

J.J. acknowledges financial
support from the National Natural Science Foundations of China through
grant 11173033
\end{acknowledgements}
\bibliographystyle{aa}
\bibliography{CER}

\end{document}